\documentclass[twocolumn,showpacs,preprintnumbers,amsmath,amssymb,citeautoscript,prb]{revtex4-1}

\usepackage{graphicx}
\usepackage{dcolumn}
\usepackage{bm}
\usepackage{times}

\begin{document}

\title{Abrupt recovery of Fermi-liquid transport by the $c$-axis collapse in CaFe$_{2}$(As$_{1-x}$P$_{x}$)$_2$ single crystals}

\author{S.~Kasahara$^{1,\ast}$}
\author{T.~Shibauchi$^{2}$}
\author{K.~Hashimoto$^{2}$}
\author{Y.~Nakai$^{2,3}$}
\author{H.~Ikeda$^{2,3}$}
\author{T.~Terashima$^{1}$}
\author{Y.~Matsuda$^{2}$}

\affiliation{$^{1}$
Research Center for Low Temperature and Materials Sciences, Kyoto University, Kyoto 606-8501, Japan
}
\affiliation{$^{2}$
Department of Physics, Kyoto University, Kyoto 606-8502, Japan
}
\affiliation{$^{3}$
TRIP, JST, Sanban-cho Building, 5, Sanban-cho, Chiyoda, Tokyo 102-0075, Japan
}

\date{\today}

\begin{abstract}
Single crystals of CaFe$_2$(As$_{1-x}$P$_x$)$_2$ are found to exhibit the tetragonal (T) to collapsed-tetragonal (cT) transition at $T_{\rm cT} \lesssim100$\,K for $x>0.05$. The $c$-axis shrinks by $\sim9$\% below $T_{\rm cT}$, which substantially diminishes the interband nesting between the hole and electron bands. In sharp contrast to the superconducting T phase of $A$Fe$_2$(As$_{1-x}$P$_x$)$_2$ ($A=$ Ba, Sr), where the anomalous non-Fermi liquid transport properties are observed, the resistivity, Hall coefficient, and magnetoresistance data in the Ca-based system all indicate that the standard Fermi liquid behaviors are recovered abruptly below $T_{\rm cT}$, and the superconductivity disappears completely. The intimate link between the superconductivity and the non-Fermi liquid transport enlightens the essential role of interband-associated fluctuation effects in Fe-pnictides. 

\end{abstract}

\maketitle


Significant deviations from the Landau's standard Fermi-liquid (FL) theory of metals are often found in strongly correlated electron systems, particularly in the vicinity of phase instabilities where an ordered state critically disappears in the zero temperature limit. Such non-Fermi liquid properties are the hallmark of strong fluctuations of the instability, which may also cause unconventional superconductivity. A prominent example is the antiferromagnetic fluctuations which are believed to govern the anomalous transport properties in e.g. heavy fermion superconductors \cite{Nakajima07}. In Fe-pnictides, the proximity between the structural and magnetic instabilities \cite{Ishida_JPSRev} leads to a more complicated situation, in which the antiferromagnetic \cite{Mazin} and/or orbital fluctuations \cite{Kontani_Onari10} may play important roles. These two kinds of fluctuations are strongly coupled with each other and they are associated with the interband quasiparticle scattering between quasi-two dimensional hole and electron Fermi-surface sheets. As these separated sheets have the same orbital character parts, the good nesting between these sheets can enhance spin fluctuations as well as orbital fluctuations. Clarifying the presence/absence of such interband fluctuations, and their connections to the superconductivity is at the core to understand the physics of Fe-pnictides. Among others, the normal-state charge transport coefficients are the most fundamental quantities involving quasiparticle masses and scattering cross sections, which can be seriously modified by these fluctuations \cite{Onari_Kontani10, Prelovsek}.

To date, however, the interpretation of the transport properties in Fe-pnictides is still under debate \cite{Rullier,Fang09,Doiron,kasa_AsP}. The deviations from the FL transport properties have been reported in several Fe-based superconductors \cite{Ishida_JPSRev,kasa_AsP,Doiron}; one of the most pronounced demonstrations is the single crystalline studies of the isovalent phosphorous substitution system BaFe$_{2}$(As$_{1-x}$P$_{x}$)$_2$ \cite{kasa_AsP, Shishido_AsP, Nakai_AsP}, which keeps the compensation condition for any value of $x$. Near the end point of the structural and antiferromagnetic transitions, the anomalous $T$-linear resistivity, $T$-dependent Hall coefficient, and the violation of Koher's rule in magnetoresistance have been reported \cite{kasa_AsP}, which are accompanied by the effective mass enhancement \cite{Shishido_AsP} as well as the NMR evidence of spin fluctuations \cite{Nakai_AsP}. However, it has been also proposed that some anomalous transport properties can be accounted for by the conventional multiband description \cite{Rullier}. To clarify this issue, it is important to study how the electronic structure (particularly interband nesting) is linked to the transport anomalies.

To this end, we find that the Ca member of the isovalent substitution series $A$Fe$_{2}$(As$_{1-x}$P$_{x}$)$_2$ ($A=$Ba, Sr, Ca) is a suitable system to study the transport properties. We have succeeded in growing single crystals of the Ca-based system, which exhibits a first-order structural phase transition from tetragonal (T) to collapsed-tetragonal (cT) phase without violations of crystalline symmetry ($I4/mmm$) for $x>0.05$. Such a structural transition with collapse of the $c$-axis parameter has been reported in the parent CaFe$_2$As$_2$ under pressure \cite{Kreyssig08, Yu09, Torikachvili_Ca, Goldman09, Canfield_review}, but had not been evident in any chemical substitution studies of CaFe$_2$As$_2$-based materials at ambient pressure including P-substituted polycrystals \cite{Kumar, Harnagea, Shi_SrCaFeAsP}. The band-structure calculations show that the Fermi surface of the hole band changes to a 3-dimensional shape, which loses the interband nesting behavior in the cT phase. In this phase we find that the anomalous transport properties are completely suppressed and conventional FL behaviors recover, and at the same time the superconductivity disappears. These results provide a direct link between the non-FL properties and the superconductivity, both of which are closely related to the interband fluctuations.

\begin{figure}[t]
\begin{center}\leavevmode
\includegraphics[width=0.95\linewidth]{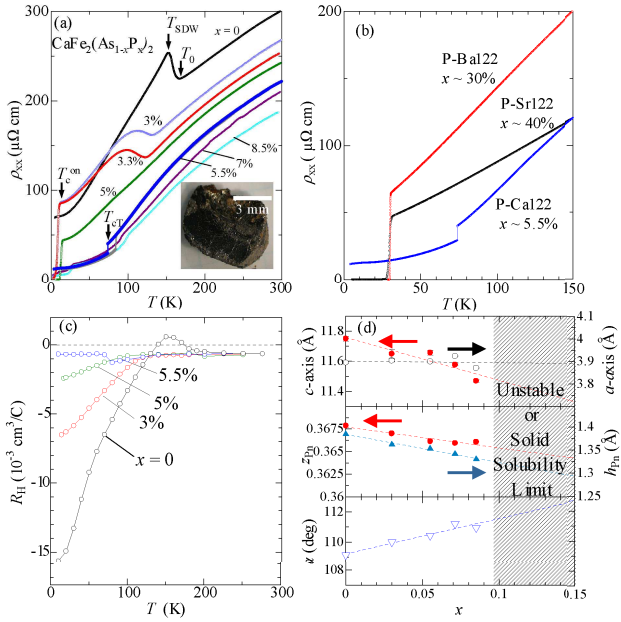}
\caption{
(Color online)
(a) $\rho_{xx}(T)$ curves of CaFe$_2$(As$_{1-x}$P$_{x}$)$_{2}$ in zero field. The inset is an optical photograph of a crystal. 
(b) Comparison of $\rho_{xx}(T)$ curves for nonmagnetic $A$Fe$_2$(As$_{1-x}$P$_x$)$_2$ ($A=$Ba, Sr, Ca) near the end point of the SDW ordered phase. 
(c) $R_H(T)$ at 5\,T 
for various $x$. 
(d) $x$-dependence of $a$- and $c$-axes lengths, the $z$ coordinate of pnictogen ($z_{\rm Pn}$) and it's height $h_{\rm Pn}$ from the 2D Fe-plane, and As-Fe-As bond angle determined by XRD. 
}
\label{RT}
\end{center}
\end{figure}

Single crystals of $A$Fe$_2$(As$_{1-x}$P$_x$)$_2$ are grown 
as described in Ref.\:\onlinecite{kasa_AsP}. 
Crystals as large as 2--5\,mm are obtained with a flat (001) surface as shown in the inset of Fig.\:\ref{RT}(a). 
As reported for polycrystalline samples \cite{Shi_SrCaFeAsP}, CaFe$_{2}$(As$_{1-x}$P$_{x}$)$_2$ crystals having higher P-concentrations ($x>0.1$) are not stable in the air even in the single crystalline form. 
$x$-values were determined by an energy dispersive X-ray analyzer. Lattice parameters are determined by single crystalline X-ray diffractions (XRD). Band structure including spin-orbit coupling is calculated by density functional theory implemented in the \textsc{Wien2k} code \cite{WIEN2k}.


Figure\:\ref{RT}(a) shows the in-plane resistivity $\rho_{xx}(T)$ of CaFe$_2$(As$_{1-x}$P$_{x}$)$_{2}$ at zero field.  
In the parent CaFe$_{2}$As$_2$, a rapid increase of $\rho_{xx}(T)$ at $T_0 = 164$ K and a peak structure at $T_{\rm SDW} = 152$ K are seen \cite{Wu_CaNa122}, which correspond to T to orthorhombic (O) structural transition and antiferromagnetic spin density wave (SDW) ordering, respectively. A tiny substitution of P for As rapidly suppresses these anomalies as a broad hump towards lower temperatures, more rapidly than the BaFe$_2$(As$_{1-x}$P$_{x}$)$_{2}$ case \cite{kasa_AsP}. 
$\rho_{xx}(T)$ shows a drop at lower temperatures and zero resistivity is attained only by $\sim 3$\% substitutions. 
The coexistence of SDW and superconductivity is observed at $x \alt 5\%$ where the anomaly is suppressed and superconductivity at $T_{\rm c} \simeq 15$ K appears. 
For the single crystals of $x \agt 5.5\%$, however, the superconducting transition disappears. 
Instead, a distinct jump in $\rho_{xx}(T)$ appears at $T_{\rm cT} \simeq 74$ K, 
accompanied by a large hysteresis between the temperature cooling and warming ($T_{\rm cT}^{\ast} \simeq 84$ K) processes, indicating the first-order nature of the transition. 
Such a jump is absent in Ba and Sr-based P-substituted systems, where the $T$-linear dependence of $\rho_{xx}(T)$ is observed at the nonmagnetic side near the end point of the SDW phase [Fig.\:\ref{RT}(b)].
For $x \agt 5.5\%$, the jump in $\rho_{xx}(T)$ becomes less pronounced and moves towards higher temperatures. 
At the same time, there also appears a noticeable reduction of $\rho_{xx}(T)$ below $\sim 30$\,K, which we attribute to the non-bulk superconductivity 
which occurs in a crystallographically separated uncollapsed T-phase likely to be present near the solid solubility limit. 
Such a phase separation origin for the partial superconductivity has been suggested for the parent CaFe$_2$As$_2$ under uniaxial pressure \cite{Torikachvili_Ca, Goldman09, Canfield_review}.

\begin{figure}[t]
\begin{center}\leavevmode
\includegraphics[width=0.93\linewidth]{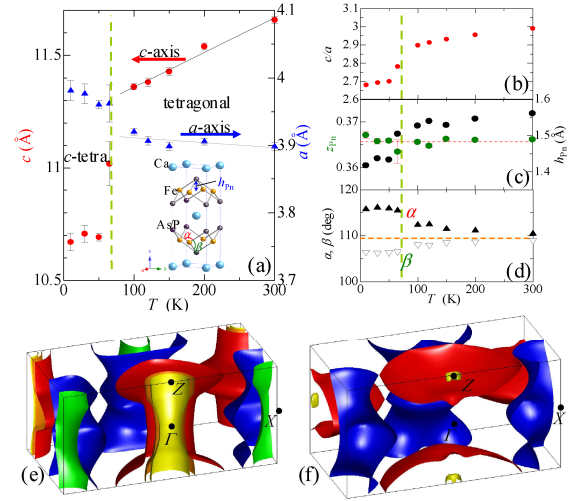}
\caption{
(Color online)
(a) Temperature dependence of $a$- and $c$-axis length of $x \approx 5.5\%$ sample, which shows \emph{pure} cT-phase. 
Inset shows schematic crystal structure. 
(b)-(d) Temperature dependence of $c/a$ ratio, 
$z_{\rm Pn}$, the pnictogen height $h_{\rm Pn}$, and the tetrahedral angle $\alpha$ ($\beta$) of the same crystal. 
The collapsed lattice constants at 10\,K are $a=3.981(15)$\,\AA, $c=10.670(38)$\,\AA, and $z_{\rm Pn} = 0.3671 (7)$. The $c$-axis reduces by $\sim9$\% and the $a$-axis expands by $\sim2$\% from room temperature values. 
(e), (f) Fermi surfaces of $x \approx 5.5$\% sample at 300 (e) and 10\,K (f) calculated by using the experimental lattice parameters. 
}
\label{XRD}
\end{center}
\end{figure}

Figure\:\ref{RT}(c) depicts the temperature dependence of Hall coefficient $R_H(T)$ measured at 5\,T. 
At low temperatures below $T_{\rm SDW}$, $R_H(T)$ shows a significant decrease. 
The low-$T$ enhancement of $|R_H(T)|$ becomes smaller as $x$ increases, but the enhancement is still there even when the system becomes non-magnetic at $x \agt 5\%$. 
At $x \approx 5.5\%$, there also appears a jump in $R_H(T)$ at $T_{\rm cT}$ as we will discuss later. 
At high temperatures above $\sim200$\,K, $R_{H}(T)$ is nearly constant and is almost independent of $x$. This is consistent with the fact that the isovalent P substitutions do not introduce charge carriers.

In Fig.\:\ref{RT}(d), we show the lattice parameters of CaFe$_2$(As$_{1-x}$P$_{x}$)$_2$ as a function of $x$ at room temperature. This demonstrates a rapid reduction of $c$-axis length whereas $a$-axis remains almost unchanged. We also see that $z$ coordinate of the pnictogen ($z_{\rm Pn}$) as well as their height $h_{\rm Pn}$ from the Fe-plane decreases as $x$ increases. The As-Fe-As bond angle $\alpha$ is enlarged correspondingly.

In Fig.\:\ref{XRD}(a), we show the temperature dependence of $a$- and $c$-axes lengths in the $x \approx 5.5\%$ crystal, which exhibits a distinct jump of $\rho_{xx}(T)$ and $R_H(T)$ with no sign of superconductivity.  
A clear reduction in the $c$-axis length as well as an expansion in the $a$-axis are observed at $T_{\rm cT}$. 
This clearly shows that a structural transition from the T to cT phase occurs in this material. 
Across the cT transition, $c/a$ ratio is significantly reduced, where $h_{\rm Pn}$ as well as $\alpha$ show distinct changes, whereas $z_{\rm Pn}$ remains almost unchanged [Fig.~\ref{XRD}(b)-(d)]. 
These lattice parameter changes are quite analogous to those found in the parent CaFe$_2$As$_2$ under pressure \cite{Kreyssig08}. 
It is also consistent with the highly uniform hydrostatic pressure studies that the \emph{pure} cT-phase does not exhibit superconductivity \cite{Yu09}.

\begin{figure}[t]
\begin{center}\leavevmode
\includegraphics[width=0.87\linewidth]{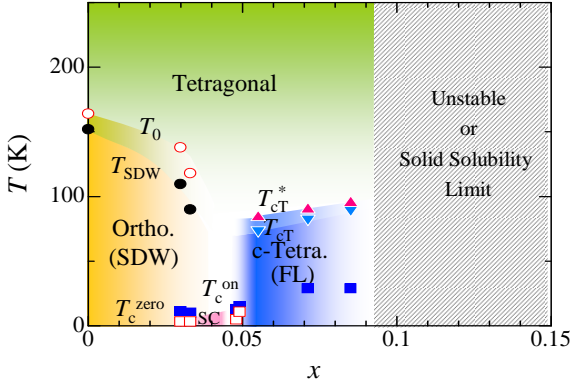}
\caption{
(Color online)
$T$--$x$ phase diagram of CaFe$_2$(As$_{1-x}$P$_{x}$)$_{2}$. The open circles show the T- to O- phase transition at $T_0$. The closed black circles show $T_{\rm SDW}$, where $\rho_{xx}(T)$ show reductions due to the reduced scattering. The solid down (up) triangles show the T to cT phase transition temperatures $T_{\rm cT}$ ($T_{\rm cT}^\ast$), where $\rho_{xx}(T)$ shows a jump in cooling (warming) processes. 
The onset of superconductivity, $T_c^{\rm on}$, and the zero resistivity temperature, $T_c^{\rm zero}$, are shown as open red and closed blue squares. 
}
\label{TvsX}
\end{center}
\end{figure}

Figure\:\ref{TvsX} demonstrates the $T$--$x$ phase diagram of CaFe$_2$(As$_{1-x}$P$_{x}$)$_2$ obtained in the present study. $T_{\rm SDW}$ and $T_0$ decrease very rapidly with $x$. At $x \alt 3.3\%$, both the SDW and superconducting transitions are visible in $\rho_{xx}(T)$. 
At $x \approx 5\%$, the superconducting phase appears below 15 K. 
Further increase of $x$ leads to an abrupt cut off of $T_c$ and the emergence of cT-phase without superconductivity. 
The structural transition at $T_{\rm cT}$ ($T_{\rm cT}^\ast$) is relatively insensitive to $x$.

Having established the phase diagram, here we focus on the impact of the T to cT phase transition. 
Figures\:\ref{XRD}(e) and (f) shows the Fermi surfaces of $x \approx 5.5\%$ crystal above and below $T_{\rm cT}$. 
In the T-phase above $T_{\rm cT}$ there are three hole sheets around the center and two electron sheets near the zone corner. 
The structural change to the cT-phase makes drastic changes in the Fermi surface structure; the two electron sheets become a single warped cylinder, whereas the hole sheets at the zone center transform to a large three-dimensional one. Such a topologically different Fermi surface structure is indeed observed in CaFe$_2$P$_2$, the end material of the present series \cite{Coldea_CaFeP}.
Obviously, the interband nesting between the hole and electron sheets is lost in the cT-phase. This is in good correspondence with the cT-phase studies of CaFe$_2$As$_2$ under pressure, which suggest the absence of magnetic fluctuations \cite{Yildrim09,Pratt_Ca,Kawasaki_Ca}.  

\begin{figure}[t]
\begin{center}\leavevmode
\includegraphics[width=0.995\linewidth]{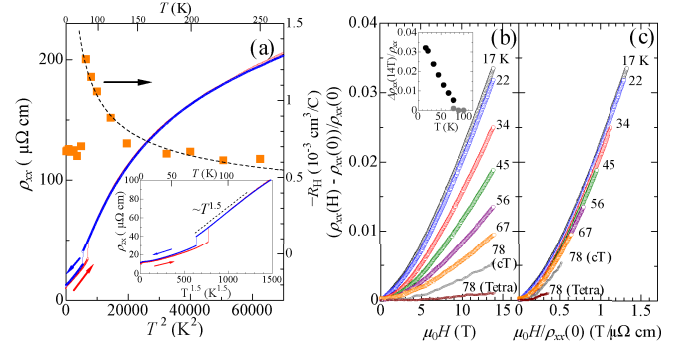}
\caption{
(Color online)
(a) $\rho_{xx}$ and $-R_{H}$ of $x \approx 5.5\%$ sample plotted against $T^2$. 
The inset shows $\rho_{xx}$ vs. $T^{1.5}$.  
(b) MR at different temperatures and (c) the Kohler's plot. Inset in (b) shows the temperature dependence of MR at 14\,T. 
}
\label{Hall}
\end{center}
\end{figure}

In Fig.\:\ref{Hall}(a) we replot $\rho_{xx}$ and $|R_H|$ against $T^2$ for the $x \approx 5.5\%$ sample. A plot of $\rho_{xx}$ vs. $T^{1.5}$ is also shown in the inset. 
In the pure cT-phase below $T_{\rm cT}$, we find that the temperature dependence of $\rho_{xx}$ is proportional to $T^2$, whereas $\rho_{xx}(T) \propto T^{1.5}$ is observed above $T_{\rm cT}$. 
The Hall coefficient $|R_H|$ is also strongly temperature dependent above $T_{\rm cT}$. 
These deviations from the standard FL properties above $T_{\rm cT}$ are similar to the behaviors found in the superconducting T-phase of Ba and Sr systems [see Fig.\:\ref{RT}(b) and Ref.\:\onlinecite{kasa_AsP}]. 
In sharp contrast, $R_H$ below $T_{\rm cT}$ is temperature independent as in the case of conventional metals. 

In the T-phase of BaFe$_2$(As$_{1-x}$P$_{x}$)$_2$ near the optimal compositions \cite{kasa_AsP}, 
it has been also observed that the magnetoresistance (MR) $\Delta\rho_{xx}(H)/\rho_{xx} \equiv [\rho_{xx}(H) - \rho_{xx}(H=0)]/\rho_{xx}(H=0)$ distinctly violates the Kohler's rule. 
Figures\:\ref{Hall}(b) and (c) demonstrate the MR data in the present Ca system. 
Except for the data at 78 K, where the system is in the hysteresis between T and cT phase, 
the different MR curves at different temperatures falls into a single curve in the cT-phase; i.e. $\Delta\rho_{xx}(H)/\rho_{xx} = f(\omega_c\tau) = f(\mu_0H/\rho_{xx}(0))$, where $f$ is a function of the cyclotron frequency $\omega_c$ and the scattering time $\tau$. This demonstrates the standard Kohler's scaling valid for conventional FL metals. 
As shown in the inset of Fig.\:\ref{Hall}(b), the magnitude of MR at 14\,T 
shows a jump from small values in the T-phase to larger values with stronger temperature dependence in the cT-phase. 
The different MR magnitudes suggest different carrier mobility in these two phases, i.e. $\tau/m$ in the T-phase is smaller than the cT-phase, most likely due to the spin/orbital fluctuations effect. 
The present results, i.e. $\rho_{xx}(T) \sim T^2$, $R_H(T) =$ const. and $\Delta\rho_{xx}(H)/\rho_{xx} = f(\omega_c\tau)$, all indicate that the charge transport properties in the cT-phase are characterized by the conventional FL behaviors. 
The negative $R_{\rm H}$ as well as the single $\omega_{c}\tau$ scaling of MR suggest that the electron contribution dominates the magneto-transport. 
The longer mean free path for electrons than for holes has been also reported by the quantum oscillation measurements in the closely related systems \cite{Shishido_AsP, Coldea_CaFeP, Analytis_AsP}.

Our results in the non-superconducting cT-phase of CaFe$_2$(As$_{1-x}$P$_{x}$)$_2$ and the comparisons with the T-phase of the related systems strongly suggest that there exists intimate relationship between the presence/absence of superconductivity and the observed non-FL/FL charge transport properties of Fe-pnictides. 
It should be noted that the non-FL transport behaviors are also observed in other strongly correlated systems, including high-$T_{c}$ cuprates and heavy fermion superconductors, which locate near antiferromagnetic instabilities. 
In those systems, it is proposed that highly anisotropic scattering time $\tau$\mbox{\boldmath $_k$} as well as the current vertex corrections due to spin fluctuations give rise to unusual non-FL transport behaviors \cite{Stojkovic97, Varma, Kontani, Nakajima07}.
In the anisotropic scattering models, the Fermi surface has hot (cold) spots, where $\tau$\mbox{\boldmath $_k$} is very short (long). 
Near the hot spots, the total current vector \mbox{\boldmath $J_{k}$} $= ne$\mbox{\boldmath $l_k$}$/\tau$\mbox{\boldmath $_k$} is no longer perpendicular to the Fermi surface, where \mbox{\boldmath $l_k$} is the mean free path vector. This is because of the excess current due to the backflow effect, which is strongly enhanced by the strong scattering between the hot spots. 
Then, $\sigma_{xy} \propto \partial\theta_k/\partial k_{\parallel}$ [$\theta_k \equiv \tan^{-1}(l_{kx}/l_{ky})$] can be enhanced, which leads to the non-FL transport properties. 

In Fe-pnictides, a recent theoretical study has suggested that anisotropic quasiparticle damping can indeed occur by the interband scattering in the presence of orbital and antiferromagnetic fluctuations \cite{Onari_Kontani10}. 
This is quite consistent with the observed recovery of FL behaviors in the cT-phase together with the reduction of interband nesting in the electronic structure. It is also important that the recovery of the FL transport is accompanied by the suppression of superconductivity, implying that such interband fluctuations are also responsible for the superconductivity in this system.

In summary, we have shown the abrupt recovery of Fermi liquid transport properties in single crystals of CaFe$_2$(As$_{1-x}$P$_{x}$)$_2$ across the tetragonal to collapsed tetragonal phase transition. 
The results strongly suggest the close relationship between the presence/absence of superconductivity and the observed non-FL/FL charge transport properties of the Fe-pnictides, which is intimately linked to the strong spin and orbital fluctuations assisted by the interband scattering.

We thank T.~Hiramatsu, Y.~Nakano, H.~Yamochi, A.~Kitada, H.~Kageyama, Y.~Maeno for technical help, and H.~Kontani, T.~Tohyama, K.~Ishida, A.~I.~Coldea, A.~Carrington for valuable discussions. 
This work is partially supported by KAKENHI and Grant-in-Aid for GCOE program 
``The Next Generation of Physics, Spun from Universality and Emergence'' from MEXT, Japan.

\end{document}